\documentclass[aps,prd,onecolumn,groupedaddress,nofootinbib,preprintnumbers,superscriptaddress]{revtex4-2}  
\usepackage{graphicx,mathtools,tabu,array}
\usepackage{epstopdf}
\usepackage{amsmath}
\usepackage{amsfonts}
\usepackage[colorlinks=true,citecolor=red,linkcolor=blue,breaklinks=true]{hyperref}
\usepackage{accents}
\usepackage[figure, table]{hypcap}
\usepackage{amssymb}
\usepackage{appendix}
\usepackage{comment}
\usepackage{bbold}
\usepackage{xcolor}
\usepackage{slashed}
\usepackage{subfigure}
\usepackage{setspace}
\usepackage{footnote}
\usepackage{multirow}
\usepackage{braket}
\usepackage[normalem]{ulem}
\usepackage[utf8]{inputenc}
\usepackage{mathrsfs}
\usepackage{longtable}
\usepackage{enumitem}
\usepackage{booktabs} 
\usepackage{subfigure}

\DeclareMathAlphabet{\mathpzc}{OT1}{pzc}{m}{it}

\graphicspath{{Figures/}}

\allowdisplaybreaks

\definecolor{palatd}{RGB}{104, 36, 109}
\definecolor{palatb}{RGB}{0, 56, 168}
\definecolor{palatr}{rgb}{0.745,0.118,0.176}
\newcommand\myshade{80}
\colorlet{mylinkcolor}{palatr}
\colorlet{mycitecolor}{palatb}
\colorlet{myurlcolor}{palatd}

\hypersetup{
  linkcolor  = mylinkcolor!\myshade!black,
  citecolor  = mycitecolor!\myshade!black,
  urlcolor   = myurlcolor!\myshade!black,
  colorlinks = true
}

\def \R  {R_{\rm CCSN}}

\begin{document}
\sloppy  

\preprint{IFT-UAM/CSIC-25-4}

\title{Dynamic Neutrino Mass Ordering and Its Imprint on \\the Diffuse Supernova Neutrino Background}

\author{Yuber F. Perez-Gonzalez}
\email{yuber.perez@uam.es}
\affiliation{Departamento de F\'{i}sica Te\'{o}rica and Instituto de F\'{i}sica Te\'{o}rica (IFT) UAM/CSIC, Universidad Aut\'{o}noma de Madrid, Cantoblanco, 28049 Madrid, Spain}

\author{Manibrata Sen}
\email{manibrata@iitb.ac.in}
\affiliation{Department of Physics, Indian Institute of Technology Bombay, Powai, Maharashtra 400076 India}

\begin{abstract}
    Neutrino masses may have evolved dynamically throughout the history of the Universe, potentially leading to a mass spectrum distinct from the normal or inverted ordering observed today. While cosmological measurements constrain the total energy density of neutrinos, they are not directly sensitive to a dynamically changing mass ordering unless future surveys achieve exceptional precision in detecting the distinct imprints of each mass eigenstate on large-scale structures. 
    In this work, we investigate the impact of a dynamic neutrino mass spectrum on the diffuse supernova neutrino background (DSNB), which is composed of neutrinos from all supernova explosions throughout cosmic history and is on the verge of experimental detection. The dynamic evolution of neutrino masses with redshift changes the propagation of neutrinos inside the supernova. Since neutrino oscillations are highly sensitive to the mass spectrum, we show that the electron neutrino survival probability carries distinct signatures of the evolving neutrino mass spectrum. Our results show that a dynamic neutrino mass spectrum can modify the DSNB flux in an energy-dependent way. However, we find that the current level of spectral shape uncertainty in DSNB modeling makes a direct detection beyond the reach of present and near-future experiments. Nonetheless, our study highlights the DSNB as a probe of redshift-dependent neutrino properties once the astrophysical systematics are brought under control.
\end{abstract}

\maketitle

\section{Introduction}

Neutrinos remain among the most elusive particles in the Standard Model (SM). Observations of solar and atmospheric neutrinos, together with results from laboratory-based neutrino oscillation experiments, have conclusively established the existence of neutrino oscillations and, consequently, non-zero neutrino masses. This discovery adds to the other unresolved problems of the SM, including the nature of dark matter and the origin of the observed matter–antimatter asymmetry in the Universe.

Neutrino oscillation experiments measure the mass-squared difference associated with the solar sector, $\Delta m^2_{\rm sol}=m_2^2-m_1^2$, as well as the absolute value of atmospheric mass-squared difference $|\Delta m^2_{\rm atm}|=|m_3^2-m_1^2|$~\cite{ParticleDataGroup:2024cfk}. Determination of the sign of $\Delta m^2_{\rm atm}$ is one of the major goals of neutrino physics. This information can be used to set limits on the sum of neutrino masses, $\sum m_i \gtrsim 0.058\,{\rm eV}$, $i=1,2,3$ in the Normal Mass Ordering (NO) or $\sum m_i \gtrsim 0.1\,{\rm eV}$ in the Inverted Mass Ordering (IO)~\cite{ParticleDataGroup:2024cfk}. Beta decay experiments like KATRIN can set an upper limit on the quantity, $\sum_i\sqrt{|U_{ei}|^2 m_i^2}< 0.45\,{\rm eV}$ ($90\% $ confidence level)~\cite{Katrin:2024tvg}, where $U$ is the Pontecorvo–Maki–Nakagawa–Sakata mixing matrix.

The tiny mass of neutrinos also plays a role in the evolution of large-scale structures in the universe and can affect cosmic background radiation, providing clues about the early universe. The sum of neutrino masses $(\sum_i m_i)$ is also a critical parameter for cosmological models, affecting dark matter and energy density estimations. In fact, observations of the Cosmic Microwave Background by PLANCK restrict the sum to be $\sum m_i<0.26\,{\rm eV}$ ~\cite{Planck:2018vyg}. Recent data on the Baryonic Acoustic Oscillations (BAO) released from the Dark Energy Spectroscopic Instrument (DESI), combined with the Cosmic Microwave Background (CMB) data from PLANCK, have set $\sum m_i< 0.072\,{\rm eV}$ with 95$\%$ C.L~\cite{DESI:2024mwx}. This stringent limit casts doubt on the IO of neutrinos, ruling it out at a significance level of around 3$\sigma$. Interestingly, the posterior probability distribution for $\sum m_i$ peaks near zero (and can go to negative values of $\sum m_i$ if positivity constraint is not imposed), \emph{suggesting potential mild inconsistencies even with the NO}\footnote{However, see~Ref.~\cite{Naredo-Tuero:2024sgf,RoyChoudhury:2024wri} for detailed discussions of the underlying assumptions in the issue of the neutrino mass going to negative values.}~\cite{Craig:2024tky,Green:2024xbb}. In fact, more recently, by integrating DESI BAO and CMB results with data from Supernova Ia, gamma-ray bursts (GRB), and X-ray observations, an even tighter constraint of $\sum m_i < 0.043\,{\rm eV}$ at the $2\sigma$ C.L. has been established, thereby confirming the inconsistency~\cite{Wang:2024hen}, see also Ref.~\cite{Jiang:2024viw} for a more comprehensive analysis.

The current experimental status on the measurement of the neutrino mass necessitates a revisit of the mechanism of neutrino mass generation, which still remains a mystery. Several theoretical models of neutrino mass have been proposed, such as the see-saw mechanisms, which introduce either heavy right-handed neutrinos, or scalar triplets or fermionic triplets in the theory~\cite{Minkowski:1977sc,Gell-Mann:1979vob,Yanagida:1979as,Mohapatra:1979ia,Mohapatra:1980yp, Schechter:1980gr}. Models with additional discrete symmetries have also been explored in great detail~\cite{King:2013eh}. The common link in the above theoretical models is that the neutrino mass arises due to a vacuum expectation value (vev) of a scalar particle (usually the Higgs). This is similar in spirit to the masses of the other SM fermions. 

However, neutrinos enjoy a special status in the SM fermion family and hence one can speculate that the neutrino mass has a totally different origin. This can either be due to any kind of novel phase transition in the early/late Universe~\cite{Fardon:2003eh,Koksbang:2017rux, Dvali:2016uhn,Lorenz:2018fzb,Lorenz:2021alz}, or due to interaction with some other particle, for e.g., dark matter~\cite{Berlin:2016woy, Krnjaic:2017zlz, Brdar:2017kbt, Capozzi:2018bps, Choi:2019zxy, Dev:2020kgz, Choi:2020ydp,Ge:2020ffj, 
Losada:2021bxx, Huang:2021kam, Chun:2021ief, Dev:2022bae, Huang:2022wmz, Davoudiasl:2023uiq, Losada:2023zap,Sen:2023uga, Sen:2024pgb,Martinez-Mirave:2024dmw,Goertz:2024gzw}. A specific feature of these mass models is that neutrino masses become redshift-dependent. Therefore, neutrino masses can be different in the early Universe from what is expected from current experiments.  This might also provide a plausible explanation if the disagreement in $\sum m_i$ between the oscillation experiments and the cosmological surveys persists. Note, however, that this does not imply that cosmological surveys will be sensitive to smaller dynamic neutrino masses. As long as the sum of neutrino masses is smaller than $\mathcal{O}(0.1)\,$eV, current cosmological surveys have no way of probing it.

It is important to emphasize at this stage that neutrino experiments can only probe the mass/mass-squared difference \emph{today}.
On the other hand, cosmological observables are actually sensitive to the neutrino energy density, which can be translated into the sum of the neutrino mass for non-relativistic neutrinos~\cite{Lesgourgues:2012uu}. So, if the individual neutrino masses, or mass-ordering, were different in the past, without violating the bound on the sum of neutrino masses from the CMB, cosmological surveys will have no way of probing it with the current/near-future sensitivities. 

This is where the Diffuse Supernova Neutrino Background (DSNB) can come in handy~\cite{Lunardini:2005jf,Beacom:2010kk}. This is composed of neutrinos coming from all possible core-collapse supernovae (SNe) from the time of star formation (redshift $\sim 6$). Neutrinos produced inside SNe propagate through the dense matter before being emitted. Depending on the value and the sign of the mass-squared differences, neutrinos may undergo a Mikheyev-Smirnov-Wolfenstein (MSW)~\cite{PhysRevD.17.2369,Mikheev:1986gs} resonant flavor conversion during propagation inside the SN. The adiabaticity of crossing the resonance also depends on the mass-squared difference and the mixing angles. A change in the mass-squared difference or the mixing angle can lead to an observable change in the DSNB spectra.

A detectable change in the DSNB spectra/flux can thus be used to probe such scenarios of dynamic neutrino mass generation~\cite{deGouvea:2022dtw}. Crucially, the expected change in the DSNB flux would not manifest as a simple normalization factor. Given that the mass generation mechanism ensures the neutrino spectrum observed today matches measurements, and it is possible that neutrinos were massless at higher redshifts, the final DSNB flux would result from a combination of SN fluxes produced under varying neutrino masses, each exhibiting distinct energy dependencies.  
Therefore, the resulting flux cannot be represented merely as a correction factor to the standard flux. Therefore, mass-varying effects can, in principle, imprint qualitatively different changes in the DSNB compared to astrophysical uncertainties.

Ref.~\cite{deGouvea:2022dtw} proposed the DSNB as a probe of late-time neutrino mass generation, focusing on scenarios in which neutrinos were effectively massless at high redshift and acquired mass at a common transition epoch. In such scenarios, the relative ordering of the neutrino mass eigenstates remains fixed, and the impact on the DSNB arises primarily from the loss of adiabatic flavor conversion inside supernovae. In this work, we extend this framework in a qualitatively new direction by allowing the individual neutrino mass eigenstates to evolve differently with redshift. This generalization permits the neutrino mass ordering itself to change over cosmic time, leading to redshift-dependent mass-squared differences whose signs may differ from their present-day values. As a result, the MSW resonance structure inside supernovae is modified in a manner that is fundamentally distinct from previously studied late-mass-generation scenarios. Our work, therefore, provides a theoretical benchmark for the type of modifications one should expect if future advances in astrophysical modeling and detector exposures reduce these uncertainties. The recent DESI results further motivate the exploration of scenarios in which the neutrino mass spectrum or ordering may have evolved over cosmic history while remaining consistent with laboratory measurements today. The framework studied in this work directly addresses this possibility by allowing for a dynamically evolving neutrino mass ordering. Therefore, the DSNB provides a unique observational window into such scenarios and complements cosmological probes that are primarily sensitive to the total neutrino energy density.

The experimental landscape of DSNB detection seems very promising. The SuperKamiokande (SK) experiment has been a front-runner in this quest. In fact, the collaboration reported a $\sim 1.5\sigma$ after analyzing the whole 5823 days of data taking before adding gadolinium~\cite{Super-Kamiokande:2021jaq}. Meanwhile, with SK doped with gadolinium, the analysis revealed a $\sim 1.2\sigma$ disagreement with a null DSNB hypothesis in its Runs VI and VII~\cite{Super-Kamiokande:2025sxh}.
Similarly, the Jiangmen Underground Neutrino Observatory (JUNO)~\cite{JUNO:2015zny,JUNO:2022lpc} has started taking data, and hopefully will present results soon.
Other promising experiments like the Hyper-Kamiokande (HK)~\cite{Hyper-Kamiokande:2018ofw,Hyper-Kamiokande:2022smq} and the Deep Underground Neutrino Experiment (DUNE)~\cite{DUNE:2020ypp,pershey2024dsnb} will join the detection efforts within the next few years. While these experiments are mostly sensitive to $\nu_e$ or $\bar{\nu}_e$, future dark matter detectors like DARWIN~\cite{DARWIN:2016hyl} will also play an important role in detecting the non-electron flavor neutrinos. Theoretically, a number of works have demonstrated the potential of the DSNB to probe new physics scenarios, which causes a spectral distortion of the DSNB~\cite{DeGouvea:2020ang,Tabrizi:2020vmo,Akita:2022etk, Das:2024ghw,Ivanez-Ballesteros:2022szu,MacDonald:2024vtw,Wang:2025qap,Roux:2024zsv,Bell:2022ycf,Ashida:2023mak,Jeong:2018yts,Farzan:2014gza,Fogli:2004gy}.
As a result, it is timely to revisit the question of whether the DSNB can shed some light on the origins of neutrino mass. 

This paper is organized as follows.
In Section~\ref{sec:dsnb}, we consider the generalities of the DSNB flux and its dependence on neutrino oscillations.
Afterwards, we consider the model for varying neutrino mass ordering as a function of redshift in Sec.~\ref{sec:model}.
We present our results in Sec.~\ref{sec:results}, and our conclusion in Sec.~\ref{sec:concl}.
We use the natural units, $\hbar = c = k_{\rm B} = 1$, throughout this manuscript.

\section{The Diffuse Supernova Neutrino Background}\label{sec:dsnb}
To predict the flux of the DSNB, it is essential to have a thorough understanding of the universe's evolution, particularly focusing on the core-collapse SN (CCSN) rate, denoted as $\R(z)$, and the neutrino flavor-dependent spectra from SNe, $F_{\nu_\beta}$. The rate of CCSN is directly tied to the star formation rate (SFR) throughout cosmic history, which has been assessed by several astronomical surveys~\cite{Hopkins:2006bw,Yuksel:2008cu,Horiuchi:2008jz}. The DSNB flux, without considering neutrino oscillations, can be expressed as,
\begin{equation}\label{eq:DSNB}
\Phi_{\nu_\beta}^0(E)=\int_0^{z_{\rm max}}\frac{dz}{H(z)}\R(z) F_{\nu_\beta}^0(E(1+z))\,,
\end{equation}
where $H(z)$ is the Hubble parameter and $E$ is the neutrino energy today, adjusted with the redshift of emission. The maximum redshift for significant star formation activity is considered to be $z_{\rm max} \approx 6$. 
We consider the CCSN rate to be related to the SFR as
\begin{equation}
    R_{\rm CCSN}(z)\approx \frac{\dot{\rho}_*(z)}{143 M_{\odot}},
\end{equation}
where $\dot{\rho}_*(z)$ is parametrized as a continuous broken power law as a function
of the redshift, see e.g.~Ref.~\cite{Beacom:2010kk},
\begin{align*}
	\dot{\rho}_*(z)=\dot{\rho}_0\left[(1+z)^{\alpha\eta}+\left(\frac{1+z}{B}\right)^{\beta\eta}+\left(\frac{1+z}{C}\right)^{\gamma\eta}\right]^{\frac{1}{\eta}}
\end{align*}
with $\dot{\rho}_0$ a normalization constant, $\eta\approx -10$, $\alpha$, $\beta$, $\gamma$ constants related to the redshift regimes, and the $B,C$ parameters related to the redshift breaks. These are given by \cite{Horiuchi:2008jz}
\begin{subequations}
    \begin{align}
		B&=(1+z_1)^{1-\frac{\alpha}{\beta}},\\
		C&=(1+z_1)^\frac{\beta-\alpha}{\gamma}(1+z_2)^{1-\frac{\beta}{\gamma}}.
    \end{align}
\end{subequations}
The parameters needed for estimating $\R(z)$ are taken from~\cite{Beacom:2010kk} and are presented in Tab.~\ref{tab:values}.
\begin{table}[t]
	\centering
	\caption{Star formation parameters used in this work, taken from \protect\cite{Horiuchi:2008jz}.}
	\label{tab:DSNBFtab}
	\begin{tabular}{c|c|c|c|c|c|c}
		\toprule
    	Fit Parameter      & $\dot{\rho}_0$ & $\alpha$ &  $\beta$ & $\gamma$ & $z_1$ & $z_2$ \\ \midrule\midrule
		Fiducial & $0.0178$ & $3.4$ & $-0.3$ &  $-3.5$ & $1$ & $4$ \\ \bottomrule
	\end{tabular}
\end{table}

The unoscillated neutrino flux from within the SN, $F_{\nu_\beta}^0(E)$, is a combination of contributions from both core-collapse SNe and black-hole-forming (BHF) failed supernovae such that 
\begin{align}
    F_{\nu_\beta}^0(E) = f_{\rm SN}\, F_{\nu_\beta}^{\rm SN}(E_\nu) + f_{\rm BH}\, F_{\nu_\beta}^{\rm BH}(E_\nu),
\end{align}
where $f_{\rm SN}$ and $f_{\rm BH}$ represent the fractions of CCSNe and BHF events, respectively~\cite{Moller:2018kpn}. We consider in what follows $f_{\rm BH} = 0.2, f_{\rm SN} = 1 - f_{\rm BH}$. These fractions can be estimated from simulations of failed SNe, and are believed to lie anywhere between $9\%$ and $41\%$~\cite{Moller:2018kpn}. We have chosen a benchmark value of $20\%$, which assumes that progenitors with masses $\gtrsim 25\,M_\odot$ form failed SNe. The functions $F_{\nu_\beta}^{\rm SN}$ and $F_{\nu_\beta}^{\rm BH}$ are the time-integrated spectra for core-collapse SNe and BHF-SNe. These spectra have been fitted using data from the Garching group simulation, specifically the results for a $12~M_\odot$ progenitor for a CCSN and a $40~M_\odot$ one for a BHF failed SNe~\cite{Garching}. This comprehensive model allows us to estimate the DSNB flux by integrating over contributions from all past SNe, giving insights into the neutrino background that is a remnant of stellar evolution across cosmic time.

There are several sources of uncertainty that plague the determination of the DSNB flux, see Ref.~\cite{Kresse:2020nto} for a compendium of different sources of uncertainty.
Although the uncertainty on the CCSN rate is generally dominant, other sources, such as the uncertain BH fraction, the neutron star mass, or the spectra shape of the emitted flux can introduce large systematic uncertainties at higher neutrino energies
State-of-the-art studies show that shape variations can reach factors of a few up near $\mathcal{O}(30)\,$MeV~\cite{Kresse:2020nto}. 
We consider the total uncertainty that includes these effects, as determined in Ref.\cite{Kresse:2020nto}, to present the standard DSNB fluxes.
We stress that our results below should be interpreted as illustrative until spectral-shape systematics are constrained by data and modeling.

Neutrinos propagating through the dense SN envelope undergo different flavor conversions. Collective oscillations, induced by neutrino-neutrino forward scattering in the dense region close to the neutrinosphere can modify the flavor composition before the onset of the MSW resonant conversion~\cite{Duan:2010bg,Mirizzi:2015eza}. However, the overall effect of collective oscillations is yet inconclusive and we expect the impact to be least in the cooling phase flux of a SN, which is what dominates the DSNB~\cite{Chakraborty:2008zp}. Hence, in this work, we neglect them for simplicity and focus on the impact of redshift-dependent mass ordering on the MSW-driven flavor evolution, assuming that the mass-squared differences remain consistent with those measured in terrestrial experiments throughout cosmic history. 
However, the electron neutrino survival probability, $P_{ee}$, may vary as a function of redshift, $z$, for non-standard scenarios, as we will explore in this work. 
Consequently, the neutrino fluxes observed at Earth must incorporate this redshift dependence.

Neglecting the effect of collective oscillations of neutrinos deep inside a SN, 
the DSNB fluxes, $\Phi_{\nu}$, can be expressed as follows
\begin{subequations}
\begin{align}\label{eq:DSNB_mvz}
\Phi_{\nu_e}(E) &= \int_0^{z_{\rm max}} \frac{dz}{H(z)} \R(z) 
\left\{ P_{ee}(z) F_{\nu_e}^0 + \left[1 - P_{ee}(z)\right] F_{\nu_x}^0 \right\}, \\
\Phi_{\bar{\nu}_e}(E) &= \int_0^{z_{\rm max}} \frac{dz}{H(z)} \R(z) 
\left\{ \overline{P_{ee}}(z) F_{\bar{\nu}_e}^0 + \left[1 - \overline{P_{ee}}(z)\right] F_{\nu_x}^0 \right\}, \\
\Phi_{\nu_x}(E) &= \int_0^{z_{\rm max}} \frac{dz}{H(z)} \R(z) 
\frac{1}{4} \left\{ 
\left[1 - P_{ee}(z)\right] F_{\nu_e}^0 + \left[1 - \overline{P_{ee}}(z)\right] F_{\bar{\nu}_e}^0 
+ \left[2 + P_{ee}(z) + \overline{P_{ee}}(z)\right] F_{\nu_x}^0 
\right\}.
\end{align}
\end{subequations}
Here, $F_{\nu_e}^0$, $F_{\bar{\nu}_e}^0$, and $F_{\nu_x}^0$ represent the initial neutrino energy spectra for electron neutrinos, electron antineutrinos, and non-electron neutrinos, respectively, while $\nu_x$ denotes all non-electronic flavors, including neutrinos and antineutrinos, i.e.,  $\nu_x\equiv\nu_\mu, \nu_\tau, \bar{\nu}_\mu, \bar{\nu}_\tau$. Since the average energy of these neutrinos within the SN is considered similar, we have not distinguished among them. The net contribution of all the non-electron flavor neutrinos to the DSNB is $4\times\Phi_{\nu_x}$.


\section{Dynamic Neutrino Mass-Squared Differences}\label{sec:model}
The premise of our study is to probe the sensitivity of the DSNB spectra to dynamic neutrino mass-squared differences in the Universe. This can arise due to redshift-dependent neutrino mass, which can be different for the different eigenstates. Such a scenario can either arise due to an unaccounted-for phase transition as the Universe evolves, or due to neutrino interactions with dynamical dark energy or ultralight dark matter. In this work, we stay agnostic of the mechanism, and assume, for demonstration purposes, that the individual neutrino masses pick up a redshift dependence as follows,
\begin{align}
    m_i = m_i^{\infty} + \frac{m_i^0 - m_i^\infty}{1+(z/z_i)^{B_i}}\,,
\end{align}
where $m_i^0$ is the mass today and $m_i^{\infty}$ is the mass at large redshifts. These two limits are joined by a smooth function, which transitions at $z=z_i$, with a rate governed by $B_i$, for each mass eigenstate $i$. 

It is crucial to emphasize that the DSNB can be sensitive to dynamic (redshift-dependent) neutrino masses, even if all the mass eigenstates transform at the same redshift, $z_s$, with the same rate, $B_s$. In that scenario, the sensitivity arises due to violation of adiabaticity in neutrino propagation as the masses become too small~\cite{deGouvea:2022dtw}. In this work, we generalize the scenario to the case where the redshift dependence is different for different eigenstates in a way that the neutrino mass ordering can be scrambled in the past. 
In this case, the mass-squared differences can become redshift-dependent and may even change sign in the past, significantly affecting neutrino flavor conversion in a highly dense environment such as a SN. 

Once neutrinos decouple and begin to free-stream from the neutrinosphere during the cooling phase of a SN explosion, they propagate through regions of extremely high density, $\rho \gg 10^{10}~{\rm g~cm^{-3}}$.  
The flavor evolution of these neutrinos is governed by the mass-squared differences and the degree to which adiabatic conditions are satisfied.  
Under adiabatic propagation, neutrinos can undergo an MSW resonant conversion upon encountering the matter resonance density, $\rho_{\rm res}$, defined as~\cite{Dighe:1999bi}  
\begin{align}
    \rho_{\rm res} = 1.4 \times 10^6~{\rm g~cm^{-3}} \left(\frac{\Delta m^2}{1~\rm eV^2}\right)\left(\frac{10~{\rm MeV}}{E_\nu}\right) \cos2\theta,
\end{align}
where $\Delta m^2$ represents the relevant mass-squared difference, $E_\nu$ is the neutrino energy, and $\theta$ is the mixing angle.  
Two primary effects arise from changes in the mass-squared differences at earlier times.  
First, the sign of the mass-squared difference determines whether the resonance occurs for neutrinos or antineutrinos.  
Second, the magnitude of the mass-squared difference influences the adiabaticity of flavor propagation. Together, these effects alter the final flavor composition of neutrinos from a SN in a distinct way, leaving an imprint on the DSNB.

\begin{figure*}
\centering
    \subfigure[$\Delta m_{21}^2 > 0, \Delta m_{31}^2 > 0$]{\includegraphics[width=0.4\textwidth]{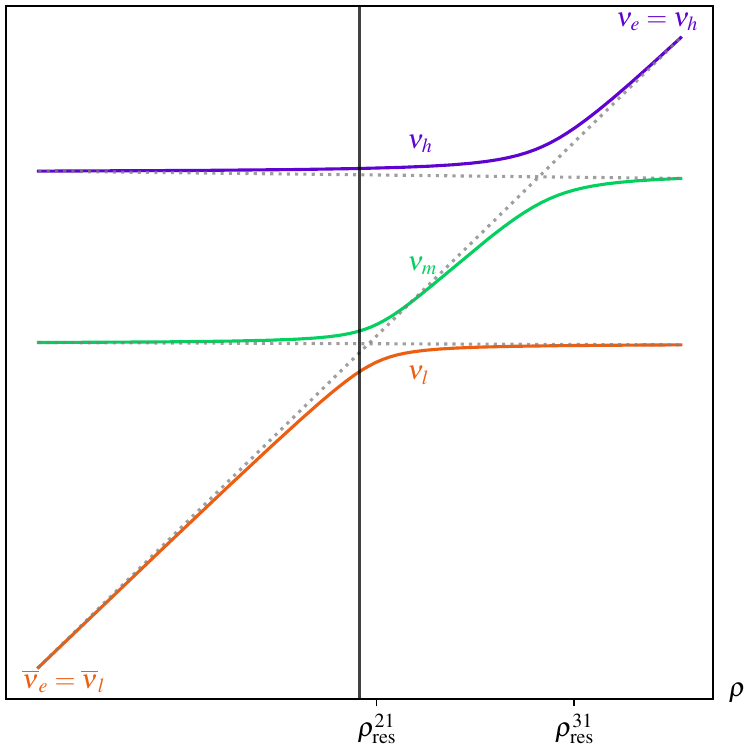}}\qquad
    \subfigure[$\Delta m_{21}^2 > 0, \Delta m_{31}^2 < 0$]{\includegraphics[width=0.4\textwidth]{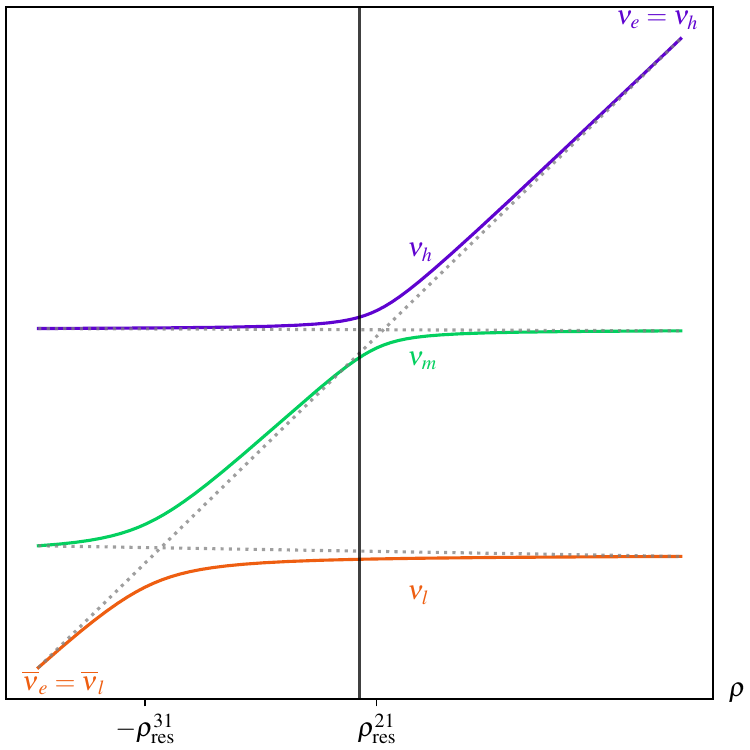}}\\
    \subfigure[$\Delta m_{21}^2 < 0, \Delta m_{31}^2 > 0$]{\includegraphics[width=0.4\textwidth]{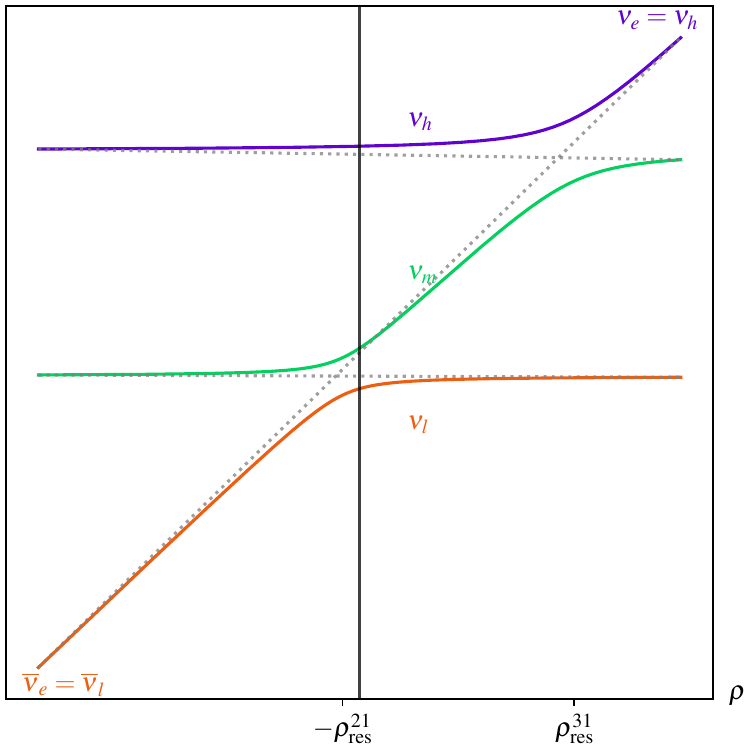}}\qquad
    \subfigure[$\Delta m_{21}^2 < 0, \Delta m_{31}^2 < 0$]{\includegraphics[width=0.4\textwidth]{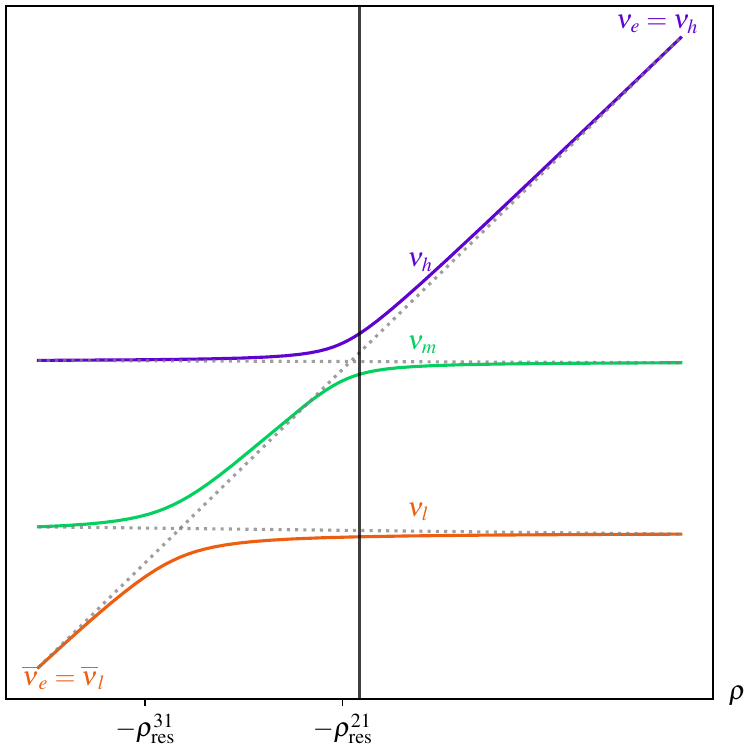}}
    \caption{Schematic depiction of level-crossing diagrams for neutrinos propagating through matter with an exponential profile for different signs of the mass splittings as a function of the density. $\nu_{h,m,l}$ correspond to the heaviest, intermediate and lightest mass eigenstates in matter, respectively. $\rho_{\rm res}^{21}$ and $\rho_{\rm res}^{31}$ refer to the values of densities that lead to a resonant enhancement of the flavor transformation for each mass splitting.\label{fig:level_cross}}
\end{figure*}

Generally, the electron neutrino survival probability can be estimated as follows.  
Inside the SN, within the high-density regions, the mixing between neutrino states becomes negligible, and the flavor states effectively coincide with the eigenstates in the medium. Specifically, for the electron flavor, we have:  
\begin{align}
    \nu_e = \nu_{h},\quad \overline{\nu}_e = \overline{\nu}_l,
\end{align} 
where \(\nu_h\) (\(\overline{\nu}_l\)) corresponds to the heaviest neutrino (lightest antineutrino) eigenstate in the medium at a given time.  
To estimate \(P_{ee}\), we analyze the level-crossing diagrams, schematically illustrated in Fig.~\ref{fig:level_cross}. These diagrams depict the eigenvalues of the Hamiltonian as functions of the matter density for a matter potential with an exponential profile, highlighting the regions where resonances occur based on the sign of the quadratic mass differences.  
As previously discussed, the occurrence of resonances for neutrinos or antineutrinos depends on the sign of the mass-squared differences. If the resonances are crossed adiabatically—that is, if the flip probability is negligible—the states will remain in their corresponding eigenstates in the medium.  
Under the assumption that adiabaticity conditions are satisfied for both resonances, whether for neutrinos or antineutrinos, the electron survival probabilities \(P_{ee}\) and \(\overline{P}_{ee}\) will be determined by the neutrino mixing matrix and can be expressed as:  
\begin{align}
    P_{ee} = |U_{eh}|^2, \quad \overline{P}_{ee} = |U_{el}|^2,
\end{align} 
where \(U_{eh}\) and \(U_{el}\) represent the elements of the mixing matrix associated with the heaviest and lightest eigenstates, respectively. In the event that adiabaticity conditions are not satisfied, one needs to take into account the flip probability - describing the hopping from one matter eigenstate to another due to the non-adiabatic nature of propagation. In the Landau-Zener formalism, the crossing probability depends on the measure of the non-adiabaticity of the propagation and is given by~\cite{Petcov:1987zj, Krastev:1988ci,Petcov:1988wv}
\begin{equation}
    P_{c} = \frac{\exp^{-(\pi\gamma F/2)} - \exp^{-(\pi\gamma F/2\sin^2\theta)}}{1-\exp^{-(\pi\gamma F/2\sin^2\theta)}},
\end{equation}
where $F$ depends on the SN matter profiles as well as the neutrino masses and mixing~\cite{Kuo:1988pn}. The parameter $\gamma$ is a measure of the non-adiabaticity of the matter profile. If the condition for MSW resonance is satisfied by the electron number density $n_e$ such that $2\sqrt{2}G_{F} E n_e = \Delta m^2\cos 2\theta$, then $\gamma$ is defined as
\begin{equation}
\gamma = \frac{\Delta m^2}{2E}\frac{\sin^2\theta}{\cos2\theta}\left(\frac{1}{n_e}\frac{dn_{e}}{dr}\right)^{-1}.
\end{equation}
Clearly, if $\gamma F >> 1$, the neutrino evolution is adiabatic. For a detailed discussion of non-adiabatic propagation and its application to our scenario, check~\cite{Kuo:1989qe, Dighe:1999bi,deGouvea:2022dtw}.

Given that the adiabaticity conditions may not have been satisfied in the Early Universe, we compute the survival probabilities of electron neutrinos and antineutrinos from a SN explosion at a given redshift $z$ numerically.  
This calculation begins with tracking the flavor evolution within the SN by numerically solving the Schrödinger-like equations, using the assumed values of the mass splittings and mixing angles at that $z$, along with the electron density profile of the presupernova star, as provided in Ref.~\cite{Garching}.  
Following this, we model the propagation of neutrinos from the SN to Earth, taking into account the redshift-dependent evolution of neutrino masses.  
This step also involves solving the Schrödinger equations, but under vacuum conditions to account for the absence of significant matter effects during interstellar travel.

\section{Results}\label{sec:results}

\begin{table}
    \centering
    \begin{tabular}{ccccccc}
    \toprule\toprule
           & $z_1$ & $z_2$ & $z_3$ & $B_1$ & $B_2$ & $B_3$ \\ \midrule\midrule
        NO & 0.500 & 0.125 & 0.050 & 9 & 6 & 3  \\ \midrule
        IO & 0.125 & 0.125 & 0.500 & 6 & 3 & 9 \\ \bottomrule
    \end{tabular}
    \caption{Assumed values for the redshift dependence of the individual mass eigenstates.}
    \label{tab:values}
\end{table}

\begin{figure*}[tb]
\centering 
    \includegraphics[width=\textwidth]{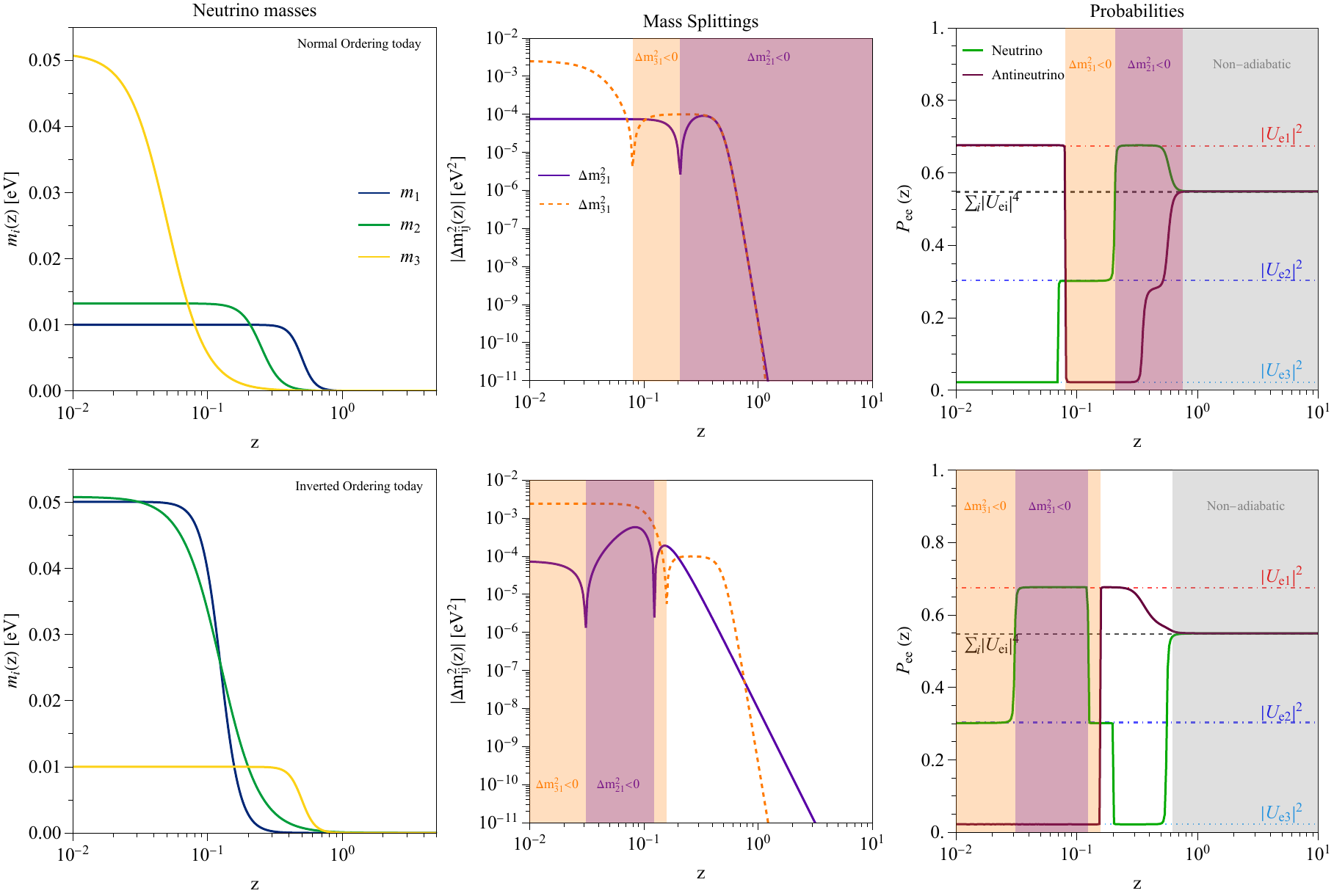}
    \caption{Varying neutrino masses and mass splittings assuming NO (upper), IO (lower) today. We plot masses (left), quadratic mass differences (center) and electron survival probabilities $P_{ee}$ at earth (right). The values used here are given in Tab.~\ref{tab:values}} \label{fig:main_plots}
\end{figure*}

We present in Fig.~\ref{fig:main_plots} the evolution of the neutrino masses (left panels), mass-squared differences (center), electron survival oscillation probabilities as measured at Earth for neutrinos and antineutrinos (right) assuming that today we have Normal (top) or Inverted (bottom) orderings.
For these figures, we have chosen the values of $z_i$ and $B_i$ as given in Tab~\ref{tab:values}.
These benchmark values are taken as illustrative cases. Changes to these parameters will shift the location of spectral features, but will not alter our qualitative results.
In the plots containing the mass splittings, we shaded the regions where the mass splittings become negative, and present in the insets the evolution of the absolute masses, with the lightest neutrino today having a mass of $m_0=0.01~{\rm eV}$.
For the probability plots, we shaded in gray the region in which the propagation inside the SN is non-adiabatic.

\begin{figure*}[tb]
\centering 
    \includegraphics[width=0.875\textwidth]{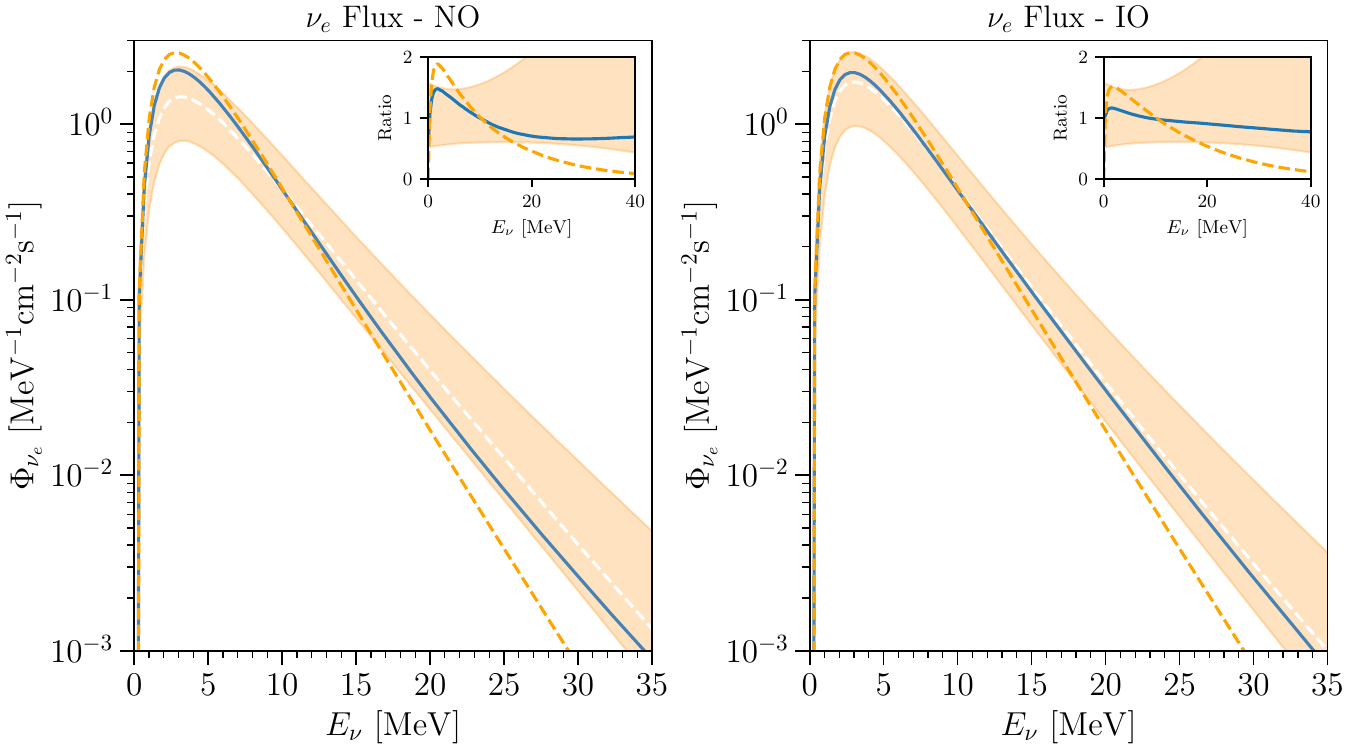}
    \caption{Illustrative modifications of the DSNB fluxes for electron neutrinos as a function of energy for NO (left) and IO (right) today for the benchmark cases presented in Tab.~\ref{tab:values} (blue) and neglecting neutrino oscillations (dashed orange). The white dashed line indicates the standard neutrino flux assuming no variation of neutrino masses, and the orange band represents the uncertainty obtained by propagating the dominant astrophysical uncertainties, including CCSN rate, failed supernova fraction, and time-integrated supernova neutrino emission spectra, following~\cite{Kresse:2020nto}. The inset indicates the ratio of the fluxes to the standard flux.}\label{fig:flux_plots}
\end{figure*}

For the NO case today, we observe the standard case for $z \lesssim 0.075$, that is, the heaviest neutrino is $\nu_3$ such that the survival probability is $|U_{e3}|^2$ for neutrinos. 
Once the neutrino masses start to change in redshift, we observe that the probability changes to $P_{ee} = |U_{e2}|^2$, as the atmospheric mass splitting becomes negative, causing the matter resonance to occur for antineutrinos rather than neutrinos. However, the solar mass splitting remains positive, enabling a MSW resonance that yields the observed probability.
This occurs in the interval $0.075 \lesssim z \lesssim 0.25$. 
For larger redshifts, $P_{ee} = |U_{e1}|^2$, as both mass splittings become negative, preventing the occurrence of resonances for neutrinos, so that neutrinos propagate adiabatically $\nu_e \leftrightarrow \nu_1$.
This will hold until the adiabaticity conditions are fulfilled, up to a value of $z\sim 0.75$. 
For larger redshifts, the neutrino masses become negligible, i.e., $m_1^\infty=
m_2^\infty = m_3^\infty \rightarrow 0$, so that the difference between the flavor and mass basis becomes unphysical, and the evolution inside the SN becomes trivial.
Thus, once neutrinos obtain their masses during their propagation to Earth, they start oscillating, and quickly decohere. The survival probability as measured at Earth then is simple
\begin{align}
    P_{ee} = \sum_i |U_{ei}|^4 \sim 0.547.
\end{align}
This result only relies on neutrino decoherence and does not depend on whether the $m_i^\infty$ are equal or not.

For antineutrinos, we observe a similar behavior, with the difference that the probability depends on whether the mass splittings become negative, enabling the MSW resonances to occur inside the SN at a given $z$, provided that adiabaticity holds.
For large redshifts $z\gtrsim 0.75$, when adiabaticity is violated, the probability tends to the same value as for neutrinos.
Finally, in the case that the ordering is inverted today, we find that the probabilities for neutrinos and antineutrinos similarly depend on which is the heaviest or lightest eigenstate when the adiabatic conditions are fulfilled, respectively. 

In Fig.~\ref{fig:flux_plots}, we present the resulting electron neutrino flux for the assumed parameters (blue), alongside the standard flux (white dashed line with orange shaded region indicating the uncertainty from different astrophysical sources, as described in Ref.~\cite{Kresse:2020nto}) and the unoscillated flux (orange dashed line). 
For neutrino energies $E_\nu \lesssim 10~\mathrm{MeV}$, the $\nu_e$ flux in the NO scenario resembles the unoscillated flux, reduced by a factor of $\sum_i |U_{ei}|^4 \sim 0.547$. This behavior is expected because the DSNB flux at these energies is dominated by neutrinos emitted at redshifts $z \gtrsim 1$. At such redshifts, the assumed mass variations imply that neutrino masses were negligible, and the primary effect is decoherence as neutrinos begin to oscillate.
At higher neutrino energies, however, the $\nu_e$ flux deviates noticeably from both the unoscillated and standard DSNB fluxes. This deviation can be attributed to the dominance of the original $\nu_x$ flux in the standard DSNB flux at Earth, which arises due to the small value of $\theta_{13}$. Since the $\nu_x$ flux has a broader energy distribution, the $\nu_e$ flux at Earth exceeds the unoscillated flux for $E_\nu \gtrsim 15~\mathrm{MeV}$.
In the mass-varying scenario, the oscillation probability is larger than $|U_{e3}|^2$, leading to a more significant contribution from the original $\nu_e$ flux at the neutrinosphere compared to the standard case. This alters the energy dependence of the flux at higher energies, and modifies its power-law behavior, although the modified flux still lies within the uncertainty region expected from different astrophysical sources. Such changes highlight the qualitative imprint of dynamic neutrino masses on the DSNB flux. Whether they can be observed depends critically on reducing the large astrophysical spectral-shape uncertainties, which at present dominate over the differences we illustrate.

Fig.~\ref{fig:flux_plots} includes an inset showing the ratio of the unoscillated and mass-varying $\nu_e$ fluxes to the standard case. This confirms the shift in the neutrino flux as a function of energy. For both the NO and the IO, we find that the modified $\nu_e$ flux lies within the astrophysical uncertainty band. The corresponding $\bar{\nu}_e$ flux does not change appreciably in the process, as explained in~\cite{deGouvea:2022dtw}. We emphasise again that while the resulting DSNB distortions may appear similar to those obtained in scenarios of late mass generation~\cite{deGouvea:2022dtw}, the two scenarios are different. In \cite{deGouvea:2022dtw}, the dominant effect originates from the loss of adiabaticity as neutrino masses vanish. In the present work, the dominant effect instead arises from redshift-dependent changes in the MSW resonance structure caused by sign changes in the mass-squared differences and a dynamically evolving mass ordering.

\begin{figure*}[tb]
\centering 
    \includegraphics[width=0.875\textwidth]{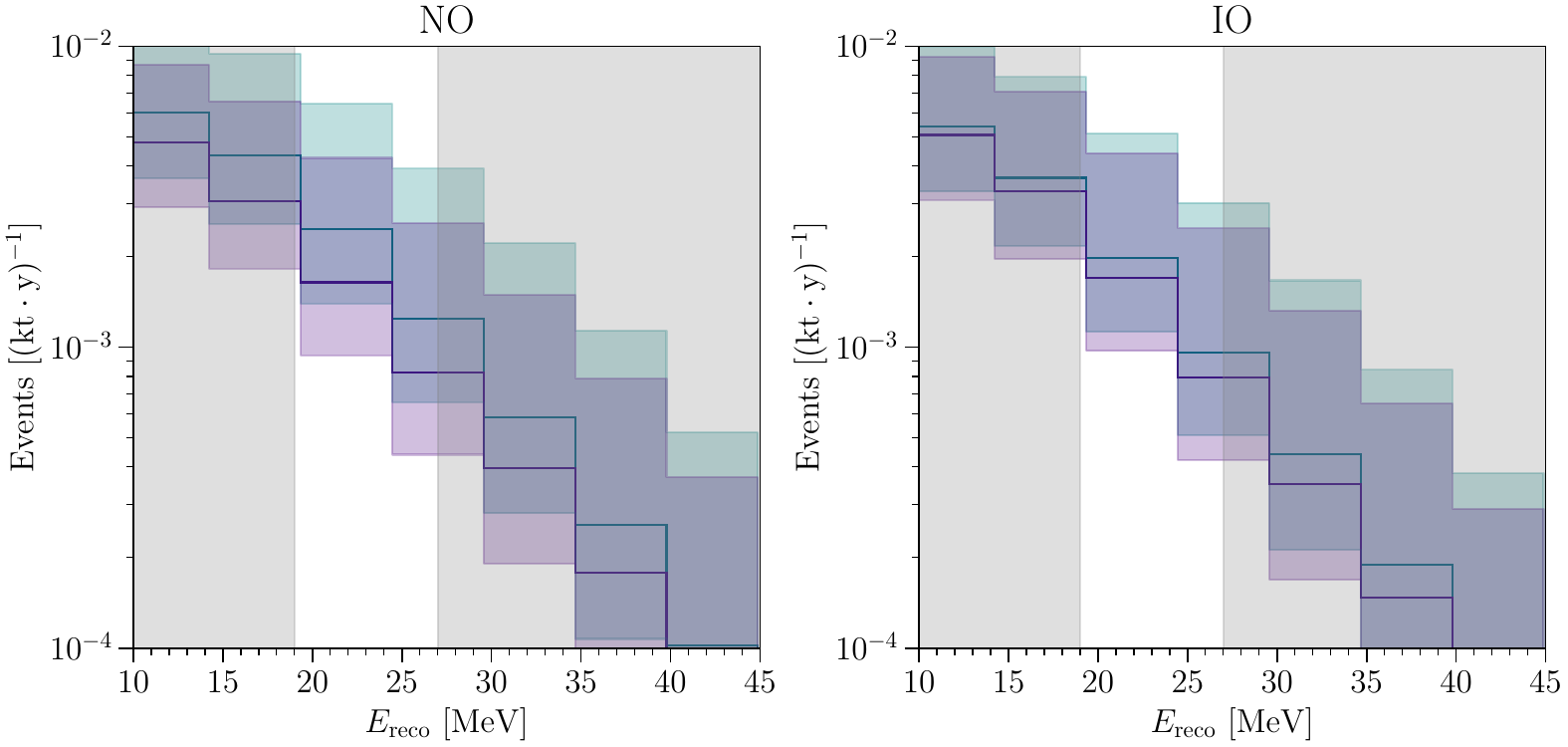}
    \caption{Expected DSNB spectra computed for a liquid argon detector as a function of the reconstructed neutrino energy. The green lines indicate the standard central values, while the green band correspond to the spectra obtained by considering the spectral shape uncertainty as discussed before. The spectra affected by the mass varying effect are plotted in purple and the values taken for this benchmark are given in Tab.~\ref{tab:values}. The gray bands indicate regions where the solar and atmospheric neutrino backgrounds significantly surpass the DSNB flux. The results are shown for fixed benchmark parameters and are not intended as sensitivity projections for DUNE.}\label{fig:spectra_plots}
\end{figure*}

To estimate the exposure required to test our mass-varying scenario, we compute the expected event spectra in a liquid argon detector via the interaction $\nu_e +\!^{40}{\rm Ar} \to e^- +\!^{40}{\rm K^*}$~\cite{Gardiner:2021qfr}, considering both the standard and mass-varying scenarios. 
Our results are shown in Fig.~\ref{fig:spectra_plots}, where the standard spectrum (green) and the mass-varying spectrum (purple) are plotted as functions of the reconstructed neutrino energy, using the parameter values listed in Tab.~\ref{tab:values}. 
The reconstructed energy is obtained by inspecting the final states in Marley after including a finite energy resolution, consistent with current measurements in liquid argon detectors~\cite{Gardiner:2021qfr,deGouvea:2020eqq}.
The green bands represent uncertainties from different astrophysical sources, as mentioned before, while the gray regions indicate energy ranges dominated by irreducible neutrino backgrounds.
Additional backgrounds are being investigated by the DUNE collaboration, but preliminary results indicate that cosmogenic and radiological backgrounds might not affect the DSNB region of interest, see e.g.~Ref.~\cite{pershey2024dsnb}.

We emphasize that the purpose of the figures is not to present a sensitivity forecast or an experimental reach, but rather to illustrate the characteristic spectral distortions induced by a dynamically evolving neutrino mass ordering. We observe that our scenario is currently degenerate with astrophysical uncertainties, the largest of them being our ignorance of the fraction of failed supernovae.
We do not perform a fit nor construct a likelihood ratio to quantify distinguishability between mass-varying scenarios and shape uncertainty due to astrophysical sources. A robust sensitivity forecast would require energy-dependent spectral-shape nuisance parameters and detector-specific backgrounds, which are beyond the scope of our study. In particular, the effects of the mass-varying scenario and the spectral shape uncertainty from the fraction of failed SNe exhibit a certain level of degeneracy. As a result, the scenarios considered here should be interpreted as theoretical benchmarks that illustrate how a dynamically evolving neutrino mass ordering would manifest in the DSNB once astrophysical and experimental uncertainties are reduced.

It is worth noting, however, that the DSNB is expected to be first discovered in the $\overline{\nu}_e$ channel, before detection in the electron neutrino channel. This is due to the ongoing efforts of large-scale detectors such as SK and HK, which are likely nearing the sensitivity needed for a first observation~\cite{Super-Kamiokande:2025sxh, beauchene2024dsnb,Santos:2025plx}. Once the $\overline{\nu}_e$ flux is measured, particularly since it is not significantly affected in the mass-varying scenario, this measurement could help constrain the overall DSNB and reduce associated uncertainties.
In this context, a clear strategy emerges for testing our scenario. Given a measurement of the $\overline{\nu}_e$ flux, one could derive the expected $\nu_e$ spectrum in both the standard and mass-varying cases and compare it with observations.

\section{Conclusion}\label{sec:concl}
The origin of the neutrino mass holds major clues to uncover the possible extensions of the Standard Model. While a number of theoretical possibilities exist, a confirmation can only be achieved after testing the predictions of the underlying models. 
A large number of terrestrial experiments have led the effort in trying to probe possible signatures arising from models of neutrino masses. On the other hand, cosmological surveys take an indirect approach and try to probe the neutrino mass through its effect on structure formation. 

This opens up the possibility of testing the dynamic origins of neutrino masses. Dynamic neutrino masses can arise due to a mechanism different from the usual mass-mechanism postulated for neutrinos. Most of the current neutrino mass models rely on the vacuum expectation value of the Higgs or any other scalar particles. On the other hand, dynamic neutrino masses can arise out of a novel phase transition, or due to dark matter-neutrino interactions, for example.
Often, in such models, neutrino masses pick up a redshift dependence and can be different in the past as compared to today. This offers an alternative mass mechanism to the usual vacuum expectation value-driven neutrino masses. However, cosmological measurements of the neutrino mass rely on the neutrino density in the early Universe to probe their masses. As a result, these surveys are not sensitive to redshift-dependent changes in the neutrino mass, as long as the energy density in the neutrinos remains the same. 

In this work, we offer a complementary probe of dynamic redshift-dependent neutrino masses in the early Universe using the Diffuse Supernova Neutrino Background (DSNB). The final flavor content of neutrinos emitted from a core-collapse supernova depends crucially on the details of neutrino propagation inside the supernova envelope. In particular, the sign of the mass-squared differences associated with the neutrinos determines whether a resonance is encountered by a neutrino or an antineutrino, and the value of the mass-squared differences determines whether the propagation is adiabatic. A combination of these two effects determines the neutrino flavor arriving at the Earth. Hence, any dynamic changes in the neutrino masses, and hence mass-orderings, will be imprinted on the DSNB. This, of course, requires that the neutrino masses change considerably within the redshift of star formation $(z\lesssim 6)$.

We demonstrated this idea by considering simple toy models of redshift-dependent neutrino masses, without alluding to any specific model. We considered the generic case where each of the neutrino masses can have a different redshift dependence. This can lead to scenarios where the mass-orderings in the past become different from what it is today. Numerically solving for the neutrino propagation inside a supernova for such different scenarios, we analyzed the differences that can arise in the DSNB. We found that the overall effect is not just a change in the normalization of the DSNB, but an alteration of the energy dependence. 
However, after computing the observable spectrum in a liquid argon detector, we found that distinguishing dynamic mass scenarios will remain challenging unless realistic DSNB spectral-shape uncertainties are tamed.
Furthermore, we emphasize that the DSNB is expected to be first detected in the $\overline{\nu}_e$ channel. In our scenario, no modifications to its spectrum are anticipated due to mass-varying effects.  
Consequently, a precise measurement of the $\overline{\nu}_e$ flux will help reduce uncertainties, potentially allowing the distinction between the standard and mass-varying scenarios with a smaller exposure.

A more general study can be performed by varying both the neutrino masses and the mixing angles with redshift to probe the dependence of the neutrino flavor on this variation. This is something we had addressed in a previous study~\cite{deGouvea:2022dtw} with other colleagues. As the mixing angle also varies and becomes smaller, we expect non-adiabatic evolution to become important. This can cause a further change in the DSNB spectra, as illustrated in~\cite{deGouvea:2022dtw}.

We would like to conclude by emphasizing the potential of the DSNB to probe the nature of the neutrino masses. If neutrino masses indeed have a dynamic nature, the only feasible way of testing this directly is through the DSNB. However, probing this effect is beyond the ability of current and future generation experiments due to the existing astrophysical uncertainties. If the uncertainties can be tamed in the future, we might have a hope of disentangling the two effects.
Current and near-future cosmological surveys will be completely insensitive to direct measurements of the neutrino mass. In this sense, our work underscores the importance of DSNB detection efforts: while current experiments are unlikely to resolve dynamic mass effects, future advances in astrophysical modeling and large-volume detectors could make the DSNB a powerful probe of new neutrino physics.

\section*{Acknowledgements}

We thank Enrique Fern\'andez-Mart\'inez, Joachim Kopp, Daniel Naredo, Ivan Mart\'inez-soler and Tim Herbermann for useful discussions. We also thank Augustine Paulaner for insightful discussions which led to the idea that shaped the manuscript. YFPG acknowledges financial support by the Consolidaci\'on Investigadora grant CNS2023-144536 from the Spanish Ministerio de Ciencia e Innovaci\'on (MCIN) and by the Spanish Research Agency (Agencia Estatal de Investigaci\'on) through the grant IFT Centro de Excelencia Severo Ochoa No CEX2020-001007-S. MS acknowledges financial support from the Centre for Advanced Study at IIT Bombay during the completion of this work.
We would also like to thank the Galileo Galilei Institute for Theoretical Physics for the hospitality and the INFN for partial support during the completion of this work.

\appendix


\bibliographystyle{apsrev4-1}
\bibliography{draft}

\end{document}